\documentclass{article}
\usepackage[utf8]{inputenc}
\usepackage[a4paper, left=25mm, right=25mm, top=20mm, bottom=20mm]{geometry}
\usepackage{mathtools}
\usepackage{physics}
\usepackage{csquotes}
\usepackage[sorting=none]{biblatex} 
\usepackage{setspace}
\usepackage{algorithm}
\usepackage{algpseudocode}
\usepackage{amsfonts}
\usepackage{scrextend}
\usepackage{verbatim}

\newtheorem{theorem}{Theorem}
\newtheorem{apptheorem}{Theorem}
\newtheorem{lemma}{Lemma}

\usepackage{float} 

\title{Classical modelling of a bosonic sampler with photon collisions}
\author{M. Umanskii, A. Rubtsov}
\date{July 2022}

\begin{document}

\maketitle

%\section{Abstract}

\abstract{
When the problem of boson sampling was first proposed, it was assumed that little or no photon collisions occur. However, modern experimental realizations rely on setups where collisions are quite common, i.e. the number of photons $M$ injected into the circuit is close to the number of detectors $N$. Here we present a classical algorithm that simulates a bosonic sampler: it calculates the probability of a given photon distribution at the interferometer outputs for a given distribution at the inputs. This algorithm is most effective in cases with multiple photon collisions, and in those cases it outperforms known algorithms.
    }

\section{Introduction}

Quantum computers are computational devices which operate using phenomena described by quantum mechanics. Therefore, they can carry out operations which are not available for classical computers. Practical tasks are known which can be solved exponentially faster using quantum computers rather than classical ones. For example, the problem of integer factorization, which underlies the widely used RSA cryptosystem, can be solved by classical computers only in exponential number of operations, whereas the quantum Shor's algorithm\cite{Shor_1997} can solve it in polynomial number of operations.

Due to technological challenges of manufacturing quantum computers, quantum supremacy (the ability of a quantum computational device to solve problems that are intractable for classical computers) in practice remains an open question. Boson sampling\cite{Lund_2014} is a good candidate for demonstrating quantum supremacy. 
Consider a linear-optics interferometer with $N$ inputs and $N$ outputs. Suppose $M$ single photons are injected into some or all of its inputs. The problem is to determine the probability distribution of states that can be observed at the outputs\cite{Gard_2015} for a given input state. Boson samplers are not universal quantum computers, that is they cannot perform arbitrary unitary rotations in the high-dimensional Hilbert space of a quantum system. Nevertheless, a simulation of a boson sampler with a classical computer requires a number of operations exponential in $M$. It was shown\cite{1011.3245} that classical complexity of boson sampling matches the complexity of computing the permanent of a complex matrix. This means that the problem of boson sampling is \#P-hard\cite{Aaronson_2011} and there are no known classical algorithms that solve it in polynomial time. The best known exact algorithm for computing the permanent of a $n \times n$ matrix is the Ryser formula\cite{RyserFormula}, which requires $O(n 2^n)$ operations. The Clifford-Clifford algorithm\cite{https://doi.org/10.48550/arxiv.1706.01260} is known to solve the boson sampling problem in $O(M 2^M + NM^2)$ operations. This makes large enough bosonic samples practically intractable with classical computational devices. Although boson sampling does not allow for arbitrary quantum computations, there are still practical problems that can be solved with boson sampling: for example, molecular docking\cite{Banchi_2020}, calculating the vibronic spectrum of a molecule\cite{Huh_2015}\cite{Huh_2017} and some graph theory problems\cite{Br_dler_2018}. Boson sampling is also useful for statistical modelling\cite{Jahangiri_2020} and machine learning\cite{PhysRevA.101.032314}\cite{Banchi_2020_1}.

There are several variants of boson sampling that aim at improving the photon generation efficiency and increasing the scale of implementations. For example, the Scattershot boson sampling uses many parametric down-conversion sources to improve the single photon generation rate. It has been implemented experimentally using a 13-mode integrated photonic chip and six PDC photon sources\cite{SBS_exp}. Another variant is Gaussian boson sampling\cite{Hamilton_2017}\cite{PhysRevLett.113.100502}, which uses Gaussian input states instead of single photons. Gaussian input states are generated using PDC sources, and it allows deterministic preparation of non-classical input light sources. In this variant, the relative input photon phases can affect the sampling distribution. Experiments were carried out with $N=12$\cite{Zhong_2019} and $N=100$\cite{Zhong_2020}. The latter implementation uses PPKTP crystals as PDC sources and employs an active phase locking mechanism to ensure coherent superposition.

%It was shown [] that a variance of the interferometer parameters of about ... changes the sampling statistics drastically, so that modelling of an ideal device makes no big sense anymore.
Any experimental set-up, of course, differs from the idealized model considered in theoretical modelling. Bosonic samplers suffer from two fundamental types of imperfections. First, the parameters of a real device, such as the reflection coefficients of the beam splitters and the phase rotations, are never known exactly. Varying the interferometer paramters too much can change the sampling statistics drastically, so that modelling of an ideal device makes no big sense anymore. Assume now that we know the parameters of the experimental set-up with great accuracy. Then what makes the device non-ideal is primarily photon losses, that is, not all photons emitted at the inputs are detected in the output channels. These losses happen because of imperfections in photon preparation, absorption inside the interferometer and imperfect detectors. There are different ways of modelling losses, for example by introducing extra beam splitters\cite{Oh_2021} or replacing the interferometer matrix   by a combination of lossless linear optics transformations and a diagonal matrix that contains transmission coefficients that are less than one\cite{Garc_a_Patr_n_2019}.

%Suppose $U$ is the $N \times N$ interferometer matrix. Then the probability of an output state will consist of the elements $U_{ij}$ of this matrix. However, this representation doesn't take into account the photon losses that are bound to happen in the experiment. 

Imperfections in  middle-sized systems make them, in general, easier to emulate with classical computers\cite{https://doi.org/10.48550/arxiv.2106.01445}. It was shown\cite{Qi_2020} that with the increase of losses in a system the complexity of the task decreases. When the number of photons $M'$ that arrive at the outputs is less than $\sqrt{M}$, the problem of boson sampling can be efficiently solved using classical computers. On the other hand, if the losses are low, the problem remains hard for classical computers\cite{Aaronson_2016}.

Photon collisions is a particular phenomenon which is present in nearly any experimental realization but was disregarded in the proposal by of Aaronson and Arkhipov\cite{1011.3245}. 
Originally it was proposed that the number of the interferometer channels is roughly a square of the number of photons in the set-up, $N \geq M^2$. In this situation, all or most of the photons arrive each to a separate channel, that is no or a few of photon collisions occur. In experimental realizations \cite{Zhong_2020}, however, $N \approx M$. %In this case the number of collisions is roughly about [?] the number of channels. 

Generally, a large number of photon collisions makes the system easier to emulate. For example, one can consider the extreme case that all photons arrive to a single output channel -- the probability of such an outcome can be estimated within a polynomial time. The effect of photon collisions on computational complexity of boson sampling has been previously studied\cite{Chin_2018}. A measure called  the Fock state concurrence sum was introduced and it was shown that minimal algorithm runtime depends on this measure. There is an algorithm for Gaussian boson sampling which also takes advantage of photon collisions\cite{doi:10.1126/sciadv.abl9236}.

In this paper, we present an algorithm aimed to simulate bosonic samplers with photon collisions. In the $N \approx M$ regime, our scheme outperforms the Clifford-Clifford method. For example, we consider an output state of the sampler with $M=N$ where one half of the outputs are empty, and the other half is populated with 2 photons in each channel.  Computing the probability of such an outcome requires us $O\left(N^2 3^{N/2} \right)$ operations. The speedup on states that have more collisions is even greater.

%We also explore the application of the Metropolis-Hastings algorithm\cite{Metropolis1953}\cite{hastings70} to the boson sampling problem. We propose a simple transition function for the Markov chain that uses the output states of the boson sampling problem as points. We show that the probability distribution of the output states calculated from the Markov chain indeed approaches the exact probability distribution, which means that the Metropolis-Hastings algorithm can be used to construct an approximate probability distribution of the output states.

\section{Problem specification}

Consider a linear-optics interferometer with $N$ inputs and $N$ outputs which is described by a given unitary $N \times N$ matrix $U$:
\begin{equation}
\tag{1}
    b_i^\dagger = \sum_{j=1}^N u_{ij} a_j^\dagger, ~ a_i^\dagger = \sum_{j=1}^N u_{ji}^* b_j^\dagger,
\end{equation}
where $a_i^\dagger$ and $b_i^\dagger$ are the creation operators on inputs and outputs respectively. We will denote the input state as 

\begin{equation}
\tag{2}
\ket{k} = \ket{k_1, k_2, ..., k_N} = \prod_{i=1}^{N} \frac{1}{\sqrt{k_i!}}(a_i^\dagger)^{k_i} \ket{0, 0, ..., 0},
\end{equation}
where $k_i$ is the number of photons in the $i$-th input. An output state will be denoted as 
\begin{equation*}
\ket{l} = \ket{l_1, l_2, ..., l_N} = \prod_{i=1}^{N} \frac{1}{\sqrt{l_i!}}(b_i^\dagger)^{l_i} \ket{0, 0, ..., 0}.
\end{equation*}

It follows from (1) and (2) that a specific input state corresponds to a set of output states that are observed with different probabilities:
\begin{equation*}
\ket{k_1, k_2, ..., k_N} = \prod_{i=1}^{N} \frac{1}{\sqrt{k_i!}}(a_i^\dagger)^{k_i} \ket{0, 0, ..., 0} =  \prod_{i=1}^{N} \frac{1}{\sqrt{k_i!}}\left(\sum_{j=1}^N u_{ji}^* b_j^\dagger\right)^{k_i} \ket{0, 0, ..., 0}
\end{equation*}

The product $\prod_{i=1}^{N} \left(\sum_{j=1}^N u_{ji}^* b_j^\dagger\right)^{k_i}$ can be written as
\begin{equation}
\tag{3}
\prod_{i=1}^{N}  \left(\sum_{j=1}^N  u_{ji}^* b_j^\dagger\right)^{k_i} =(u_{11}^* b_1^\dagger + u_{21}^* b_2^\dagger + ... + u_{N1}^* b_{N}^\dagger)^{k_1}  \cdot  ... \cdot (u_{1N}^* b_1^\dagger + u_{2N}^* b_2^\dagger + ... + u_{NN}^* b_{N}^\dagger)^{k_N}.
\end{equation}

After expanding, this expression will be a sum of terms that have the following form:
\begin{equation*}
\alpha(l_1, l_2, ..., l_N) \prod_{i=1}^{N} \frac{1}{\sqrt{k_i!}} \left(b_i^\dagger\right)^{l_i} \ket{0, 0, ..., 0} = \alpha(l_1, l_2, ..., l_N) \prod_{i=1}^N \sqrt{\frac{l_i!}{k_i!}} \ket{l_1, l_2, ..., l_N}, 
\end{equation*}
where $\alpha(l_1, l_2, ..., l_N)$ is a complex number that consists of the elements of $U$ that correspond to the given output state. Therefore, the probability of observing an output state $\ket{l_1, l_2, ..., l_N}$ will be
\begin{equation*}
\left|\braket{l_1, l_2, ..., l_N}{ k_1, k_2, ..., k_N}\right|^2 = \left| \alpha(l_1, l_2, ..., l_N) \prod_{i=1}^N \sqrt{\frac{l_i!}{k_i!}} \right|^2 = \left|\alpha(l_1, l_2, ..., l_N)\right|^2 \prod_{i=1}^N \frac{l_i!}{k_i!}.
\end{equation*}
The problem consists in determining the probabilities of all of output states. The main difficulty lies in calculating the number $\alpha(l_1, l_2, ..., l_N)$ for given input and output states. This paper presents an algorithm that solves this problem using the properties of the Fourier transform.

\section{Algorithm description}

Let us define a function
\begin{equation}
\tag{4}
g(t; \{Q_i\}) = \prod_{p=1}^{N} \left( \sum_{q=1}^{N} e^{i 2 \pi Q_q t} ~u_{qp}^*  \right)^{k_p}, 
\end{equation}
where $\{Q_i\}$ is some fixed set of $N$ natural numbers. The choice of $\{Q_i\}$ will later be discussed in detail. This function represents the expression (3), where creation operators $b_j^\dagger$ are replaced with exponents $e^{i 2 \pi Q_j t}$ that oscillate with frequencies $Q_j$.

After expanding the expression (4), we get the following:
\begin{equation*}
    g(t; \{Q_i\}) = \prod_{p=1}^{N} \left( \sum_{q=1}^{N} e^{i 2 \pi Q_q t} ~u_{qp}^* \right)^{k_p} = 
\end{equation*}
\begin{equation*}
= (e^{i 2 \pi Q_1 t} u_{11}^* + e^{i 2 \pi Q_2 t} u_{21}^* + ...+ e^{i 2 \pi Q_{N} t} u_{N1}^* )^{k_1} \cdot  ... \cdot (e^{i 2 \pi Q_1 t} u_{1N}^* + e^{i 2 \pi Q_2 t} u_{2N}^* + ... + e^{i 2 \pi Q_{N} t} u_{NN}^* )^{k_N} =
\end{equation*}
\begin{equation*}
= \sum e^{i 2 \pi \sum_{i=1}^{N}l_i Q_i t} ~\alpha(l_1, l_2, ..., l_{N}),
\end{equation*}
where the sum is computed over all sets $\{{l_1, ..., l_{N}}\}$ such that $\sum_{i=1}^{N} l_i = M$.

Therefore, for each possible output state $\ket{l} = \ket{l_1, l_2, ..., l_N}$ there is a harmonic in $g(t; \{Q_i\})$ that has a frequency of $f(\ket{l}; \{Q_i\}) = \sum_{i=1}^{N} l_i Q_i$ and an amplitude of $\alpha(l_1, l_2, ..., l_{N})$. The set of numbers $\{Q_i\}$ can be chosen in such a way that the harmonics don't overlap, i.e. there are no two outputs states $\ket{l}$ and $\ket{l'}$ with equal frequencies $f(\ket{l}; \{Q_i\}) = f(\ket{l'}; \{Q_i\})$.

If no harmonics overlap, then any of the numbers $\alpha(l_1, l_2, ..., l_{N})$ can be found from the Fourier transform of the function $g(t; \{Q_i\})$. On the other hand, to calculate the probability of a specific state $\ket{l}$ it is sufficient to choose $\{Q_i\}$ in such a way that the frequency $f(\ket{l}; \{Q_i\})$ is unique in the spectrum, i.e. the frequency of any other state $\ket{l'}$ differs from the frequency of the state in consideration: $f(\ket{l}; \{Q_i\}) \neq f(\ket{l'}; \{Q_i\}) ~ \forall \ket{l'} \neq \ket{l}$.

An example of $g(t, \{Q_i\})$ with non-overlapping harmonics and its spectrum can be seen in Figure 1 (the choice of $\{Q_i\}$ used here is described in section 3.1).
\begin{figure}[H]
\centering
\includegraphics[scale=0.5]{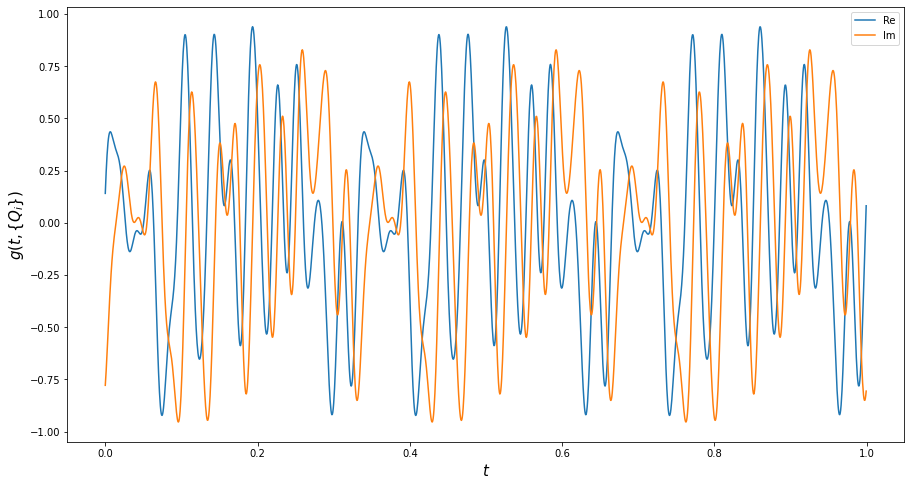}
\includegraphics[scale=0.5]{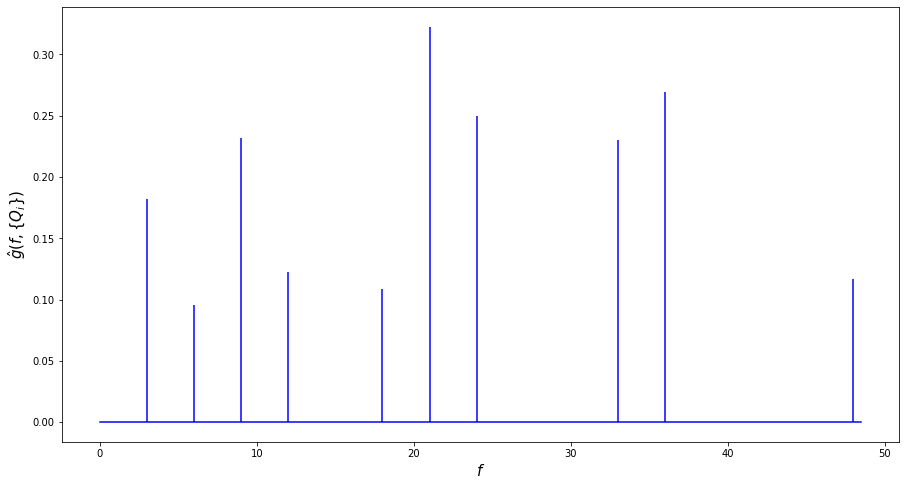}
\caption{An example of the function $g(t, \{Q_i\})$ (the top picture) and its spectrum (the bottom picture). A system with $N=M=3$ is used, the input state is $\ket{k} = \ket{1,1,1}$. Each peak in the spectrum corresponds to one of ten possible output states.}
\end{figure}

\subsection{The first method of choosing $\{Q_i\}$}

Let us consider the methods of choosing $\{Q_i\}$ that will satisfy the necessary conditions on the spectrum. The first one consists in the following: let $M$ be the total number of photons at the inputs, i.e. $M=\sum_{i=1}^{N} k_i$ for an input state $\ket{k_1, k_2, ..., k_N}$. We choose $Q = \{ 1, M+1, (M+1)^2, ..., (M+1)^{N-1} \}$, or $Q_i = (M+1)^{i-1}$. Then for any output state $\ket{l_1, l_2, ..., l_N}$ the sum
\begin{equation*}
   \sum_{i=1}^{N} l_i Q_i = 1 \cdot l_1 + (M+1) \cdot l_2 + (M+1)^2 \cdot l_3 + ... + (M+1)^{N-1} \cdot l_N
\end{equation*}
will be a number that has a representation $\overline{l_Nl_{N-1}...l_1}$ in a positional numeral system with radix $M+1$ (since $l_i < M+1 ~ \forall i$). From the uniqueness of representation of numbers in positional numeral systems it follows that every sum $\sum_{i=1}^{N} l_i Q_i$ (some number in a positional numeral system with radix $M+1$) will correspond to exactly one set of numbers $l_1, l_2, ..., l_N$ (its representation in this numeral system; $l_i$ being its digits).

Using this method of choosing $\{Q_i\}$ guarantees that the probability of any output state can be calculated from the spectrum of $g(t; \{Q_i\})$, since the frequencies $f(\ket{l}; \{Q_i\})$, $f(\ket{l'}; \{Q_i\})$ are different for any two output states $\ket{l} \neq \ket{l'}$.

\subsection{The second method of choosing $\{Q_i\}$}

Another method of choosing $\{Q_i\}$ is useful when the goal is to compute the probability of one specific output state $\ket{l}$ when the input state $\ket{k}$ is given. This method doesn't guarantee that the frequencies will be different for any two output states, but it guarantees that the frequency of the state in consideration (the target frequency) $f(\ket{l}; \{Q_i\})$ will be unique in the spectrum. Note that in this case $Q_i = Q_i(\ket{l})$, i.e. the choice of $\{Q_i\}$ depends on the output state.

This method of choosing $\{Q_i\}$ can be described in the following way:
\begin{equation}
\tag{5}
    Q_i = \begin{cases}
        \prod_{j=1}^{i-1} (l_j+1), ~l_i \neq 0 \\
        0, ~ l_i = 0;
    \end{cases}
\end{equation}
\begin{equation*}
    Q_1 = \begin{cases}
        1, ~l_1 \neq 0 \\
        0, ~ l_1 = 0.
    \end{cases}
\end{equation*}

Therefore if all of the outputs in the state $\ket{l}$ contain photons, then $Q_1 = 1$, $Q_2 = (l_1+1)$, $Q_3 = (l_2+1)(l_1+1)$ and so on: $Q_{i+1}$ is $(l_i+1)$ times greater than $Q_i$. 

Let us show that this method will actually lead to the target frequency being unique in the spectrum. Let $\{ Q_i \} = \{ Q_i(\ket{l}) \}$ be the frequencies calculated using the method described above. We need to prove that for any output state $\ket{l'} \neq \ket{l}$ it is true that $f(\ket{l}; \{Q_i\}) \neq f(\ket{l'}; \{Q_i\})$, i.e.
\begin{equation*}
    \sum_{i=1}^N l_i Q_i \neq \sum_{i=1}^N l'_i Q_i.
\end{equation*}

Firstly, let's suppose that some of the outputs in the state $\ket{l_1, l_2, ..., l_N}$ contain $0$ photons. Let $h_1, ... , h_K$ be the indices of the outputs that contain at least one photon: $l_{h_i} > 0 ~ \forall i \in \{1, 2, ..., K\}; ~~ K<N$. Then the condition  $f(\ket{l}; \{Q_i\}) \neq f(\ket{l'}; \{Q_i\})$ becomes
\begin{equation*}
\sum_{i=1}^{K} l_{h_i} Q_{h_i} \neq \sum_{i=1}^{K} l'_{h_i} Q_{h_i},
\end{equation*}
since all the terms corresponding to empty outputs are zero in both sums ($Q_i=0$ if the $i$-th output contains $0$ photons).

Note that we can view it as a "reduced"\ system with $K$ outputs, in which the output state $\ket{l}$ contains at least one photon in every output. However, this system has one difference. Previously we considered the possible output states to be all states that satisfy $\sum_{i=1}^N l'_i = M$ and $l'_i \geq 0 ~ \forall i$. Now, in this "reduced"\ system we must consider all output states such that $\sum_{i=1}^K l'_i \leq M$ and $l'_i \geq 0 ~ \forall i$. This happens because output states $\ket{l'}$ can have a non-zero amount of photons in outputs that were empty in  $\ket{l}$; such outputs will have no effect on the frequency and they remain outside the "reduced"\ system.

Therefore, instead of a system where some outputs can be empty and some $Q_i$ can be zero, but $\sum_{i=1}^N l_i = \sum_{i=1}^N l'_i$, we can consider a system where $l_i > 0 ~ \forall i$ but  $\sum_{i=1}^N l_i \geq \sum_{i=1}^N l'_i$. In this system $Q_i = \prod_{j=1}^{i-1} (l_j+1)$, and $Q_1 = 1$.

To prove the correctness of the algorithm, we must prove the following statement:
\begin{theorem}
Let $N$ be some natural number. Let $l_1, l_2, ..., l_N, l'_1, l'_2, ... l'_N$ be natural numbers that satisfy the following conditions: \\
1) $l_i > 0,~ l'_i \geq 0 ~\forall i$;\\
2) $\sum_{i=1}^{N}l_i \geq \sum_{i=1}^N l'_i$;\\
3) $\sum_{i=1}^N l_i Q_i = \sum_{i=1}^N l'_i Q_i$, where $Q_i = \prod_{j=1}^{i-1} (l_j+1)$ (and $Q_1 = 1$). \\
Then $l_i = l'_i ~ \forall i$.
\end{theorem}

The proof of this statement can be found in the Appendix.

\section{Parameters of the Fourier transform}

To calculate the Fourier transform of the function $g(t; \{Q_i\})$ we will use a fast Fourier transform (FFT). Firstly, we will define its parameters: the sampling interval $\Delta t$ (or the sampling frequency $f_s = \frac{1}{\Delta t}$) and the number of data points $K$. The function will be calculated at points $n\Delta t, ~ 1 \leq n \leq K$. Since all the frequencies in the spectrum of $g(t; \{Q_i\})$ are natural numbers, they can be discerned with the frequency resolution of $\Delta f = 1$. The function therefore will be calculated in points within an interval $[0;1]$ which contains at least one period of each of the harmonics.

The sampling frequency $f_s = \frac{1}{\Delta t}$ is often chosen according to the Nyquist-Shannon theorem: if the Nyquist frequency $f_N = \frac{f_s}{2} = \frac{1}{2 \Delta t}$ is greater than the highest frequency in the spectrum $f_{max}$, then the function can be reconstructed from the spectrum and no aliasing occurs. Therefore, one way of choosing the sampling frequency is $f_s = 2 f_{max}$. It can be used with both methods of choosing $\{Q_i\}$. 

Since the function is calculated in points within an interval $[0;1]$, the number of data points $K$ is equal to the sampling frequency $f_s$. Optimization of the algorithm requires lowering the sampling frequency as much as possible.

If the goal is to calculate the probability of one specific state $\ket{l}$ and the second method of choosing $\{Q_i\}$ is used, then the sampling frequency $f_s$ can be chosen to be lower than $2 f_{max}$. This will lead to aliasing: a peak with frequency $f$ will be aliased by peaks with frequencies $f + k f_s, ~ k \in \mathbb {Z}$. To correctly calculate the probability of the output state from the spectrum computed this way, the spectrum must not contain frequencies that satisfy $f(\ket{l'}; \{Q_i\}) = f(\ket{l}; \{Q_i\}) + k f_s, ~ k \in \mathbb {Z}$. Note that it won't be possible to reconstruct the function $g(t; \{Q_i\})$ from such a spectrum.

We will show that the sampling frequency $f_s$ for calculating the probability of an output state $\ket{l}$ using the second method of choosing $\{Q_i\}$ can be chosen to be greater by one than the target frequency $f=f(\ket{l}; \{Q_i(\ket{l})\})$:
\begin{equation*}
    f_s = f + 1 = \sum_{i=1}^{N} l_i Q_i + 1 
\end{equation*}

To prove this statement, we must show that the spectrum of $g(t; \{Q_i\})$ won't contain any frequencies  $f' = f(\ket{l'}; \{Q_i\})$ that satisfy $f' = f + k f_s, ~ k \in \mathbb {Z}$. This is shown by a theorem that is analogous to Theorem 1 yet has a weaker condition: equation in condition 3) is taken modulo $f_s$.

\begin{theorem}
Let $N$ be some natural number. Let $l_1, l_2, ..., l_N, l'_1, l'_2, ... l'_N$ be natural numbers that satisfy the following conditions: \\
1) $l_i > 0,~ l'_i \geq 0 ~\forall i$;\\
2) $\sum_{i=1}^{N}l_i \geq \sum_{i=1}^N l'_i$;\\
3) $\sum_{i=1}^N l_i Q_i \equiv \sum_{i=1}^N l'_i Q_i \mod{( \sum_{i=1}^{N} l_i Q_i + 1)}$, where $Q_i = \prod_{j=1}^{i-1} (l_j+1)$ (and $Q_1 = 1$). \\
Then $l_i = l'_i ~ \forall i$.
\end{theorem}

The proof of this statement can be found in the appendix.

\section{Complexity of the algorithm}

Let's consider the computational complexity of this algorithm. The complexity of a fast Fourier transform on a data array of $K$ points is $O(K \log{K})$. Total complexity of the algorithm consists of the complexity of calculating $g(t; \{Q_i\})$ in $K$ points and the complexity of a fast Fourier transform.

Computing $g(t; \{Q_i\})$ in each point is done in $\propto N^2$ operations: the expression for $g(t; \{Q_i\})$ consists of at most $N$ factors, each of which can be computed in $N$ additions, $N$ multiplications and $N$ exponentiations. If some of the inputs are empty, there will be fewer factors in the expression, and the resulting complexity will be lower.

When the first method of choosing $\{Q_i\}$ is used, the number of data points $K$ is proportional to the highest frequency in the spectrum of $g(t; \{Q_i\})$, since the sample frequency is chosen using the Nyquist-Shannon theorem. The frequencies corresponding to the outputs states in this case are equal to $\sum_{i=1}^{N} l_i Q_i = \sum_{i=1}^{N} l_i (M+1)^{i-1}$. The highest frequency then is $M(M+1)^{N-1}$ and corresponds to the state where the last output contains all the photons. The total complexity of calculating all the probabilities then will be
\begin{equation*}
O \left( N^2 M (M+1)^{N-1}\right) + O\left(M(M+1)^{N-1} \log{\left(M(M+1)^{N-1}\right)} \right) =
\end{equation*}
\begin{equation*}
    = O \left( N^2 M^N + N M^N \log{M} \right).
\end{equation*}

When the second method of choosing $\{Q_i\}$ is used, the number of data points $K$ depends on the frequency of the output state in consideration. This frequency is highest when photons are spread over outputs evenly. For a system with $M = mN$, $m \in \mathbb{N}$ this corresponds to a state where each output contains $m$ photons. In this case the highest frequency is equal to 
\begin{equation*}
\sum_{i=1}^N m (m+1)^{i-1} = m \frac{(m+1)^{N-1} - 1}{m} = (m+1)^{N-1} - 1.
\end{equation*}

Therefore, the sampling frequency and the required number of data points will be $(m+1)^{N-1}$. The complexity of the algorithm in the worst case will be
\begin{equation*}
    O\left(N^2 (m+1)^N + N(m+1)^N \log{(m+1)}  \right).
\end{equation*}

In most states, however, photons won't be spread evenly between outputs, and outputs with high number of photons will lower the sampling frequency and the complexity for calculating the probability of the state. This means that the more photon collisions are in a state, the better this algorithm performs.
Let's consider several specific cases.

1. $M=N$, the goal is to compute the output state that contains $2$ photons in one half of the outputs and $0$ photons in the other half. The frequency corresponding to such state will be
\begin{equation*}
\sum_{i=1}^{N/2} 2 \cdot 3^{i-1} = 3^{N/2-1} - 1, 
\end{equation*}
and the complexity of the algorithm will be equal to
\begin{equation*}
O\left(\frac{N^2}{2} (3^{N/2-1} - 1) + (3^{N/2-1} - 1)\log{(3^{N/2-1} - 1)} \right) = O\left(N^2 3^{N/2} + N 3^{N/2} \right) = O\left(N^2 3^{N/2} \right).
\end{equation*}

%FIXME: 2^N + N^3 -> 2^N
For comparison, the complexity of the Clifford-Clifford algorithm (which is $O(M 2^M + NM^2)$) in this case will be equal to $O(N 2^N)$.

2. $M=N^2$, the goal is to compute the output state that contains $2N$ photons in one half of the outputs and $0$ photons in the other half. The frequency corresponding to such state will be $\sum_{i=1}^{N/2} 2N \cdot (2N+1)^{i-1} = (2N+1)^{N/2-1} - 1$, and the algorithm complexity will be 
\begin{equation*}
    O\left(N^2 (2N)^{N/2} + N (2N)^{N/2} \log{N} \right) = O\left(N^2 \cdot (2N)^{N/2} \right)
\end{equation*}

Again, the complexity of the Clifford-Clifford algorithm in this case will be equal to $O(N^2 2^{N^2})$.

\subsection{Weighted average complexity}

We can measure the weighted average computational complexity of the algorithm described above by computing $\sum_i p_i C_i$, where $p_i$ is the probability of the $i$-th state, $C_i$ is the complexity of calculating the probability of $i$-th state (assuming the second method of choosing $\{Q_i\}$ is used), and the sum is calculated over all possible states.

We have computed this weighted average complexity for systems with varying $N$. We set $M=N$, and $\ket{k}=\ket{1,1,...,1}$ as the input state. The interferometer matrices for those systems were randomly generated unitary matrices. Figure 2 shows that the weighted average complexity of the algorithm is significantly lower than the maximum complexity of the algorithm and their ratio decreases as $N$ increases.

\begin{figure}[H]
\centering
\includegraphics[scale=0.5]{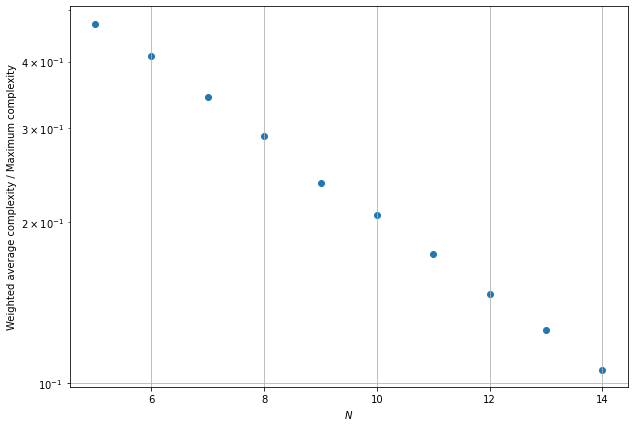}
\caption{Decrease of the ratio of weighted average complexity to maximum complexity with the increase of $N$.}
\end{figure}

\section{The Metropolis-Hastings algorithm}

For systems with large $N$ it might be computationally intractable to calculate the exact probability distribution of output states. The number of possible output states scales with $N$ and $M$ as
\begin{equation*}
C_{M+N-1}^{N-1} = \frac{(M+N-1)!}{(N-1)!(M+N-1-(N-1))!} = \frac{(M+N-1)!}{(N-1)!M!}.
\end{equation*}
Sampling from a probability distribution from which direct sampling is difficult can be done using Metropolis-Hastings algorithm, which uses a Markov process. It allows to generate a Markov chain in which points appear with frequencies that are equal to their probability. In our case, the points will be represented by the output states, i.e. sets of numbers $\ket{l_1, l_2, ..., l_N}$ such that $\sum_{i=1}^N l_i = M$.

We will require a transition function that will generate a candidate state from the last state in the chain. When the points are represented by real numbers, a candidate state can be chosen from a Gaussian distribution centered at the last point. In our case, however, the transition function will be more complex.

The transition function must allow the chain to arrive in each of the possible states. It will be convenient to define it in the following way: 

\begin{algorithm}
\caption{Transition function}\label{alg:cap}
\begin{algorithmic}
\State $h := \{ i: ~ l_i > 0 \}$ \Comment{$h$ is a set that contains indices of non-empty outputs}
\State $K := |h|$ \Comment{$K$ is the amount of non-empty outputs}
\State $r := random(\{1, ..., K\})$ \Comment{We generate a random number $1 \leq r \leq K$}
\State $l_{h_{r}} := l_{h_{r}} - 1$ \Comment{Decrease the number of photons in $h_{r}$-th output}
\State $s := random(\{1, ..., N\} \setminus {h_{r}})$  \Comment{We generate a random number $1 \leq s \leq N$ such that $s \neq r$}
\State $l_{s} := l_{s} + 1$ \Comment{Increase the number of photons in $s$-th output}
\State return $\ket{l_1, l_2, ..., l_N}$
\end{algorithmic}
\end{algorithm}

For the condition of detailed balance to hold, we will require a function $p(l_1, ..., l_N; l'_1, ..., l'_N)$ which is equal to the probability of $\ket{l'_1, ..., l'_N}$ being the transition function output when the last state in the chain is $\ket{l_1, ..., l_N}$. It is trivially constructed from the transition function.

Let $u$ be the ratio of the exact probabilities of states $\ket{l'_1, ..., l'_N}$ and $\ket{l_1, ..., l_N}$. The condition of detailed balance will hold if the Markov chain will go from state $\ket{l_1, ..., l_N}$ to state $\ket{l'_1, ..., l'_N}$ with the probability 
\begin{equation*}
    \alpha = u \cdot \frac{p(l'_1, ..., l'_N; l_1, ..., l_N)}{p(l_1, ..., l_N; l'_1, ..., l'_N)}.
\end{equation*}

Given the Markov chain, we can then calculate the approximate probability of a state by dividing the number of times this state occurs in the chain by the total number of steps of the chain.

\subsection{Results}

To demonstrate that the frequencies with which states appear in the Markov chain converge to the exact probability distribution, we have tested it on a system with $N=10, ~M=10, ~\ket{k}=\ket{1,1,...,1}$ and a random unitary $10 \times 10$ matrix as the interferometer matrix. To calculate the distance between the exact and the approximate distribution we used cosine similarity:
\begin{equation*}
S_C(p,q) = \frac{(\overrightarrow{p} \cdot \overrightarrow{q})}{|\overrightarrow{p}| \cdot |\overrightarrow{q}|} = \frac{\sum_{i} p_i q_i}{\sqrt{\left(\sum_i p_i^2 \right)}\cdot \sqrt{\left(\sum_i q_i^2 \right)}},
\end{equation*}
where $p$ and $q$ are some probability distributions. Namely, the value of $1-S_C(p,q)$ is $0$ when $p$ and $q$ are equal.

Figure 3 shows that $1-S_C(p,q)$ decreases as the Markov chain makes more steps.
\begin{figure}[H]
\centering
\includegraphics[scale=0.5]{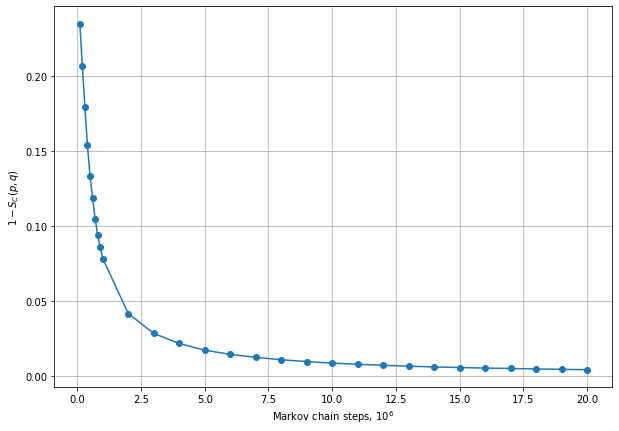}
\caption{Convergence of the approximate probability distribution to the exact probability distribution.}
\end{figure}

\section{Conclusion}

We have presented a new algorithm for calculating the probabilities of the output states in the boson sampling problem. We have shown the correctness of the algorithm and calculated its computational complexity. This algorithm is simple in implementation as it relies heavily on the Fourier transform, which has numerous well-documented implementations. 

The performance of this algorithm is better than the other algorithms in cases where there are many photon collisions. An example we give is an output state where all the photons are spread equally across one half of the outputs, with the other half of the outputs empty. In this case the algorithm requires $O\left(N^2 3^{N/2} \right)$ operations, while the Clifford-Clifford algorithm requires $O\left(N 2^N\right)$ operations.

We have also proposed a method to approximately calculate the probability distribution in the boson sampling problem. It can be used when the system size is too large and calculating the exact probability distribution is intractable. Our results show that this algorithm indeed produces a probability distribution that converges to the exact probability distribution.

We plan to study further the application of the Metropolis-Hastings algorithm to approximating the boson sampling problem. When losses are modelled in the system, the probability distribution of the output states becomes concentrated. For example, when losses are high, the most probable states are those with many lost photons. When the losses are low, the probability is concentrated in the area where no or a few photons are lost. This property makes the Metropolis-Hastings algorithm especially effective in solving this problem.

\printbibliography

@online{1011.3245,
Author = {Scott Aaronson and Alex Arkhipov},
Title = {The Computational Complexity of Linear Optics},
Year = {2010},
Eprint = {1011.3245},
Eprinttype = {arXiv},
}

@misc{https://doi.org/10.48550/arxiv.2106.01445,
  doi = {10.48550/ARXIV.2106.01445},
  url = {https://arxiv.org/abs/2106.01445},
  author = {Popova, A. S. and Rubtsov, A. N.},
  title = {Cracking the Quantum Advantage threshold for Gaussian Boson Sampling},
  publisher = {arXiv},
  year = {2021},
  copyright = {arXiv.org perpetual, non-exclusive license}
}

@article{Garc_a_Patr_n_2019,
	doi = {10.22331/q-2019-08-05-169},
	url = {https://doi.org/10.22331%2Fq-2019-08-05-169},
	year = 2019,
	month = {aug},
	publisher = {Verein zur Forderung des Open Access Publizierens in den Quantenwissenschaften},
	volume = {3},
	pages = {169},
	author = {Raúl García-Patrón and Jelmer J. Renema and Valery Shchesnovich},
	title = {Simulating boson sampling in lossy architectures},
	journal = {Quantum}
}

@article{Oh_2021,
	doi = {10.1103/physreva.104.022407},
	url = {https://doi.org/10.1103%2Fphysreva.104.022407},
	year = 2021,
	month = {aug},
	publisher = {American Physical Society ({APS})},
	volume = {104},
	number = {2},
	author = {Changhun Oh and Kyungjoo Noh and Bill Fefferman and Liang Jiang},
	title = {Classical simulation of lossy boson sampling using matrix product operators},
	journal = {Physical Review A}
}

@misc{https://doi.org/10.48550/arxiv.1706.01260,
  doi = {10.48550/ARXIV.1706.01260},
  url = {https://arxiv.org/abs/1706.01260},
  author = {Clifford, Peter and Clifford, Raphaël},
  title = {The Classical Complexity of Boson Sampling},
  publisher = {arXiv},
  year = {2017},
  copyright = {arXiv.org perpetual, non-exclusive license}
}

@incollection{Gard_2015,
	doi = {10.1142/9789814678704_0008},
	url = {https://doi.org/10.1142%2F9789814678704_0008},
	year = 2015,
	month = {jun},
	publisher = {{WORLD} {SCIENTIFIC}
},
	pages = {167--192},
	author = {Bryan T. Gard and Keith R. Motes and Jonathan P. Olson and Peter P. Rohde and Jonathan P. Dowling},
	title = {An Introduction to Boson-Sampling},
	booktitle = {From Atomic to Mesoscale}
}

@article{Lund_2014,
	doi = {10.1103/physrevlett.113.100502},
	url = {https://doi.org/10.1103%2Fphysrevlett.113.100502},
	year = 2014,
	month = {sep},
	publisher = {American Physical Society ({APS})},
	volume = {113},
	number = {10},
	author = {A.{\hspace{0.167em}
}P. Lund and A. Laing and S. Rahimi-Keshari and T. Rudolph and J.{\hspace{0.167em}}L. O'Brien and T.{\hspace{0.167em}}C. Ralph},
	title = {Boson Sampling from a Gaussian State},
	journal = {Physical Review Letters}
}

@article{Huh_2015,
	doi = {10.1038/nphoton.2015.153},
	url = {https://doi.org/10.1038%2Fnphoton.2015.153},
	year = 2015,
	month = {aug},
	publisher = {Springer Science and Business Media {LLC}
},
	volume = {9},
	number = {9},
	pages = {615--620},
	author = {Joonsuk Huh and Gian Giacomo Guerreschi and Borja Peropadre and Jarrod R. McClean and Al{\'{a}}n Aspuru-Guzik},
	title = {Boson sampling for molecular vibronic spectra},
	journal = {Nature Photonics}
}

@article{Huh_2017,
	doi = {10.1038/s41598-017-07770-z},
	url = {https://doi.org/10.1038%2Fs41598-017-07770-z},
	year = 2017,
	month = {aug},
	publisher = {Springer Science and Business Media {LLC}
},
	volume = {7},
	number = {1},
	author = {Joonsuk Huh and Man-Hong Yung},
	title = {Vibronic Boson Sampling: Generalized Gaussian Boson Sampling for Molecular Vibronic Spectra at Finite Temperature},
	journal = {Scientific Reports}
}

@article{Banchi_2020,
	doi = {10.1126/sciadv.aax1950},
	url = {https://doi.org/10.1126%2Fsciadv.aax1950},
	year = 2020,
	month = {jun},
	publisher = {American Association for the Advancement of Science ({AAAS})},
	volume = {6},
	number = {23},
	author = {Leonardo Banchi and Mark Fingerhuth and Tomas Babej and Christopher Ing and Juan Miguel Arrazola},
	title = {Molecular docking with Gaussian Boson Sampling},
	journal = {Science Advances}
}

@article{Jahangiri_2020,
	doi = {10.1103/physreve.101.022134},
	url = {https://doi.org/10.1103%2Fphysreve.101.022134},
	year = 2020,
	month = {feb},
	publisher = {American Physical Society ({APS})},
	volume = {101},
	number = {2},
	author = {Soran Jahangiri and Juan Miguel Arrazola and Nicol{\'{a}
}s Quesada and Nathan Killoran},
	title = {Point processes with Gaussian boson sampling},
	journal = {Physical Review E}
}

@article{Br_dler_2018,
	doi = {10.1103/physreva.98.032310},
	url = {https://doi.org/10.1103%2Fphysreva.98.032310},
	year = 2018,
	month = {sep},
	publisher = {American Physical Society ({APS})},
	volume = {98},
	number = {3},
	author = {Kamil Brádler and Pierre-Luc Dallaire-Demers and Patrick Rebentrost and Daiqin Su and Christian Weedbrook},
	title = {Gaussian boson sampling for perfect matchings of arbitrary graphs},
	journal = {Physical Review A}
}

@article{PhysRevA.101.032314,
  title = {Measuring the similarity of graphs with a Gaussian boson sampler},
  author = {Schuld, Maria and Br\'adler, Kamil and Israel, Robert and Su, Daiqin and Gupt, Brajesh},
  journal = {Phys. Rev. A},
  volume = {101},
  issue = {3},
  pages = {032314},
  numpages = {11},
  year = {2020},
  month = {Mar},
  publisher = {American Physical Society},
  doi = {10.1103/PhysRevA.101.032314},
  url = {https://link.aps.org/doi/10.1103/PhysRevA.101.032314}
}

@article{Banchi_2020_1,
	doi = {10.1103/physreva.102.012417},
	url = {https://doi.org/10.1103%2Fphysreva.102.012417},
	year = 2020,
	month = {jul},
	publisher = {American Physical Society ({APS})},
	volume = {102},
	number = {1},
	author = {Leonardo Banchi and Nicol{\'{a}
}s Quesada and Juan Miguel Arrazola},
	title = {Training Gaussian boson sampling distributions},
	journal = {Physical Review A}
}

@article{Shor_1997,
	doi = {10.1137/s0097539795293172},
	url = {https://doi.org/10.1137%2Fs0097539795293172},
	year = 1997,
	month = {oct},
	publisher = {Society for Industrial {\&} Applied Mathematics ({SIAM})},
	volume = {26},
	number = {5},
	pages = {1484--1509},
	author = {Peter W. Shor},
	title = {Polynomial-Time Algorithms for Prime Factorization and Discrete Logarithms on a Quantum Computer},
	journal = {{SIAM} Journal on Computing}
}

@article{Aaronson_2011,
	doi = {10.1098/rspa.2011.0232},
	url = {https://doi.org/10.1098%2Frspa.2011.0232},
	year = 2011,
	month = {jul},
	publisher = {The Royal Society},
	volume = {467},
	number = {2136},
	pages = {3393--3405},
	author = {Scott Aaronson},
	title = {A linear-optical proof that the permanent is \#P-hard},
	journal = {Proceedings of the Royal Society A: Mathematical, Physical and Engineering Sciences}
}

@article{Qi_2020,
	doi = {10.1103/physrevlett.124.100502},
	url = {https://doi.org/10.1103%2Fphysrevlett.124.100502},
	year = 2020,
	month = {mar},
	publisher = {American Physical Society ({APS})},
	volume = {124},
	number = {10},
	author = {Haoyu Qi and Daniel J. Brod and Nicol{\'{a}
}s Quesada and Raúl García-Patrón},
	title = {Regimes of Classical Simulability for Noisy Gaussian Boson Sampling},
	journal = {Physical Review Letters}
}

@article{Aaronson_2016,
	doi = {10.1103/physreva.93.012335},
	url = {https://doi.org/10.1103%2Fphysreva.93.012335},
	year = 2016,
	month = {jan},
	publisher = {American Physical Society ({APS})},
	volume = {93},
	number = {1},
	author = {Scott Aaronson and Daniel J. Brod},
	title = {{BosonSampling} with lost photons},
	journal = {Physical Review A}
}

@article{Zhong_2020,
	doi = {10.1126/science.abe8770},
	url = {https://doi.org/10.1126%2Fscience.abe8770},
	year = 2020,
	month = {dec},
	publisher = {American Association for the Advancement of Science ({AAAS})},
	volume = {370},
	number = {6523},
	pages = {1460--1463},
	author = {Han-Sen Zhong and Hui Wang and Yu-Hao Deng and Ming-Cheng Chen and Li-Chao Peng and Yi-Han Luo and Jian Qin and Dian Wu and Xing Ding and Yi Hu and Peng Hu and Xiao-Yan Yang and Wei-Jun Zhang and Hao Li and Yuxuan Li and Xiao Jiang and Lin Gan and Guangwen Yang and Lixing You and Zhen Wang and Li Li and Nai-Le Liu and Chao-Yang Lu and Jian-Wei Pan},
	title = {Quantum computational advantage using photons},
	journal = {Science}
}

@article{PhysRevLett.113.100502,
  title = {Boson Sampling from a Gaussian State},
  author = {Lund, A. P. and Laing, A. and Rahimi-Keshari, S. and Rudolph, T. and O'Brien, J. L. and Ralph, T. C.},
  journal = {Phys. Rev. Lett.},
  volume = {113},
  issue = {10},
  pages = {100502},
  numpages = {5},
  year = {2014},
  month = {Sep},
  publisher = {American Physical Society},
  doi = {10.1103/PhysRevLett.113.100502},
  url = {https://link.aps.org/doi/10.1103/PhysRevLett.113.100502}
}

@article{doi:10.1126/sciadv.abl9236,
author = {Jacob F. F. Bulmer  and Bryn A. Bell  and Rachel S. Chadwick  and Alex E. Jones  and Diana Moise  and Alessandro Rigazzi  and Jan Thorbecke  and Utz-Uwe Haus  and Thomas Van Vaerenbergh  and Raj B. Patel  and Ian A. Walmsley  and Anthony Laing },
title = {The boundary for quantum advantage in Gaussian boson sampling},
journal = {Science Advances},
volume = {8},
number = {4},
pages = {eabl9236},
year = {2022},
doi = {10.1126/sciadv.abl9236},
URL = {https://www.science.org/doi/abs/10.1126/sciadv.abl9236},
eprint = {https://www.science.org/doi/pdf/10.1126/sciadv.abl9236}}

@article{Chin_2018,
	doi = {10.1038/s41598-018-24302-5},
	url = {https://doi.org/10.1038%2Fs41598-018-24302-5},
	year = 2018,
	month = {apr},
	publisher = {Springer Science and Business Media {LLC}
},
	volume = {8},
	number = {1},
	author = {Seungbeom Chin and Joonsuk Huh},
	title = {Generalized concurrence in boson sampling},
	journal = {Scientific Reports}
}

@book{RyserFormula,
title = {Combinatorial Mathematics},
journal = {Carus Mathematical Monograph},
volume = {14},
year = {1963},
author = {Herbert John Ryser},
}

@article{SBS_exp,
doi = 10.1126/sciadv.1400255,
author = {Marco Bentivegna  and Nicolò Spagnolo  and Chiara Vitelli  and Fulvio Flamini  and Niko Viggianiello  and Ludovico Latmiral  and Paolo Mataloni  and Daniel J. Brod  and Ernesto F. Galvão  and Andrea Crespi  and Roberta Ramponi  and Roberto Osellame  and Fabio Sciarrino },
title = {Experimental scattershot boson sampling},
journal = {Science Advances},
volume = {1},
number = {3},
pages = {e1400255},
year = {2015},
doi = {10.1126/sciadv.1400255},
URL = {https://www.science.org/doi/abs/10.1126/sciadv.1400255},
eprint = {https://www.science.org/doi/pdf/10.1126/sciadv.1400255}}

@article{Hamilton_2017,
	doi = {10.1103/physrevlett.119.170501},
	url = {https://doi.org/10.1103%2Fphysrevlett.119.170501},
	year = 2017,
	month = {oct},
	publisher = {American Physical Society ({APS})},
	volume = {119},
	number = {17},
	author = {Craig S. Hamilton and Regina Kruse and Linda Sansoni and Sonja Barkhofen and Christine Silberhorn and Igor Jex},
	title = {Gaussian Boson Sampling},
	journal = {Physical Review Letters}
}

@article{Zhong_2019,
	doi = {10.1016/j.scib.2019.04.007},
    url = {https://doi.org/10.1016%2Fj.scib.2019.04.007},
	year = 2019,
	month = {apr},
	publisher = {Elsevier {BV}
},
	volume = {64},
	number = {8},
	pages = {511--515},
	author = {Han-Sen Zhong and Li-Chao Peng and Yuan Li and Yi Hu and Wei Li and Jian Qin and Dian Wu and Weijun Zhang and Hao Li and Lu Zhang and Zhen Wang and Lixing You and Xiao Jiang and Li Li and Nai-Le Liu and Jonathan P. Dowling and Chao-Yang Lu and Jian-Wei Pan},
	title = {Experimental Gaussian Boson sampling},
	journal = {Science Bulletin}
}

\section{Appendix}

\begin{apptheorem}
Let $N$ be some natural number. Let $l_1, l_2, ..., l_N, l'_1, l'_2, ... l'_N$ be natural numbers that satisfy the following conditions: \\
1) $l_i > 0,~ l'_i \geq 0 ~\forall i$;\\
2) $\sum_{i=1}^{N}l_i \geq \sum_{i=1}^N l'_i$;\\
3) $\sum_{i=1}^N l_i Q_i = \sum_{i=1}^N l'_i Q_i$, where $Q_i = \prod_{j=1}^{i-1} (l_j+1)$ (and $Q_1 = 1$). \\
Then $l_i = l'_i ~ \forall i$.
\end{apptheorem}
\emph{Proof.} We will prove this theorem by induction on $N$. The base case will be $N=2$. Both the base case and the induction step will be proven by contradiction.

1. Base case.

Let's assume the opposite: $l_1 \neq l'_1$ and/or $l_2 \neq l'_2$.

The condition 3) will take the form

\begin{equation*}
l_1 Q_1 + l_2 Q_2 = l'_1 Q_1 + l'_2 Q_2 
\end{equation*}
\begin{equation*}
l_1+l_2(l_1+1) = l'_1+l'_2(l_1+1) .
\end{equation*}

We expand the brackets:

\begin{equation*}
l_1+l_2 + l_2 l_1 = l'_1+l'_2 +  l'_2 l_1.
\end{equation*}

According to condition 2), $l_1+l_2 \geq l'_1 + l'_2$. Therefore,

\begin{equation*}
l_2 l_1 \leq l'_2 l_1.
\end{equation*}

Condition 1) states that $l_1>0$. Therefore, $l_2 \leq l'_2$.

On the other hand, let us write condition 3) modulo $(l_1+1)$; the terms containing $(l_1+1)$ will be zero:

\begin{equation*}
    l_1 \equiv l'_1 \mod{(l_1+1)}
\end{equation*}

$l_1 \neq l'_1$, since otherwise it follows from condition 3) that $l_2(l_1+1) = l'_2(l_1+1) \implies l_2 = l'_2$ which leads to a contradiction (both $l_1=l'_1$ and $l_2=l'_2$). Therefore, since $l_1 \geq 0$ and $l'_1 \geq 0$, we have $l_2 > l'_2$. However, previously we have shown that $l_2 \leq l'_2$, which leads to a contradiction. This proves the base case.

2. Let us prove some general statements that will help us prove the induction step. Let's assume the statement of the theorem is true for $N-1$. Then a following lemma holds for numbers  $l_1,...,l_N,l'_1,...,l'_N$ that satisfy the conditions of the theorem for $N$:

\begin{lemma}
$\forall m \in \mathbb {N}: 1 \leq m \leq N$ the following is true: $l'_m = l_m - k_{m-1} + k_m(l_m+1)$, where $k_m$ and $k_{m-1}$ are natural numbers and $k_0=k_N=0$. 
\end{lemma}
\emph{Proof.} We will prove this lemma by induction on $m$. First we prove the base case $m=1$. Let's write down the expression from condition 3) of the theorem modulo $(l_1+1)$ (all terms that contain $(l_1+1)$ will turn to zero and only the first ones from each side will remain):

\begin{equation}
    \tag{A.1}
    l_1 \equiv l'_1 \mod{(l_1+1)}.
\end{equation}

Then 

\begin{equation}
    \tag{A.2}
    l'_1 = l_1 + k_1(l_1+1),
\end{equation}

where $k_1$ is an integer. Since $l'_1 \geq 0$, $k_1$ must be natural. Since $k_0 = 0$, the equation $l'_1 = l_1 - k_0 + k_1(l_1+1)$ is true.

Now let's assume that the statement of this lemma holds for all $i$ such that $1 \leq i \leq m-1$. Let's write down the expression from condition 3) of the theorem modulo $(l_{m}+1)(l_{m-1}+1)...(l_1+1)$ - this will turn to zero all the terms, except for first $m$ on both sides (note that $Q_m$ has a form of $(l_{m-1}+1)(l_{m-2}+1)...(l_1+1)$):

\begin{equation*}
    l_1 Q_1 + l_2 Q_2+...+l_m Q_m  \equiv  l'_1 Q_1 + l'_2 Q_2 + ... + l'_m Q_m \mod{(l_{m}+1)(l_{m-1}+1)...(l_1+1)}.
\end{equation*}

Then 

\begin{equation}
\tag{A.3}
l'_1 Q_1 + l'_2 Q_2+...+l'_m Q_m = l_1 Q_1 + l_2 Q_2 +...+l_m Q_m + k_m(l_m+1)(l_{m-1}+1)...(l_1+1),
\end{equation}

where $k_m$ is an integer. Using $l'_1 = l_1 + k_1(l_1+1)$ and dividing by $(l_1+1)$ , we get

\begin{equation*}
k_1 + l'_2 + l'_3(l_2+1)+...+l'_m(l_{m-1}+1)(l_{m-2}+1)...(l_2+1) =
\end{equation*}
\begin{equation*}
= l_2 + l_3(l_2+1)+...+l_m(l_{m-1}+1)(l_{m-2}+1)...(l_2+1) + k_m(l_m+1)(l_{m-1}+1)...(l_2+1).
\end{equation*}

Sequentially applying $l'_i = l_i - k_{i-1} + k_i(l_i+1)$ and dividing by $(l_i+1)$ for all $i$ from $2$ to $m-1$ we get the following:

\begin{equation*}
k_{m-1} + l'_m = l_m + k_m (l_m+1) \implies l'_m = l_m - k_{m-1} + k_m(l_m+1),
\end{equation*}

where $k_{m-1} \geq 0$ by the assumption of the induction step and $k_m \geq 0$ since $l'_m \geq 0$.

Let's specifically consider the case $m=N$. The expression (A.3) will take the following form:

\begin{equation*}
l_1 Q_1 + l_2 Q_2+...+l_N Q_N  =  l'_1 Q_1 + l'_2 Q_2 + ... + l'_N Q_N + k_N(l_N+1)(l_{N-1}+1)...(l_1+1).
\end{equation*}

Using condition 3) of the theorem we get 

\begin{equation*}
k_N(l_N+1)(l_{N-1}+1)...(l_1+1) = 0 \implies k_N = 0,
\end{equation*}

which proves the lemma.

We now go back to the theorem. Suppose the theorem is false for $N$ but true for $N-1$. Lemma 1 has some corollaries that are used in proving the induction step. Firstly, suppose $l_1=l'_1$ but $\exists i>1 : l_i\neq l'_i$. Then condition 3) will take the form of

\begin{multline*}
l_2(l_1+1)+l_3(l_2+1)(l_1+1)+...+l_{N}(l_{N-1}+1)...(l_1+1)= \\
=l'_2(l_1+1)+l'_3(l_2+1)(l_1+1)+...+l'_{N}(l_{N-1}+1)...(l_1+1),
\end{multline*}

which can be divided by $(l_1+1)$ to get

\begin{equation*}
l_2+l_3(l_2+1)+...+l_{N}(l_{N-1}+1)...(l_2+1)=l'_2+l'_3(l_2+1)+...+l'_{N}(l_{N-1}+1)...(l_2+1).
\end{equation*}

Moreover, condition 2) can be written as

\begin{equation*}
\sum_{i=2}^{N}l_i \geq \sum_{i=2}^{N} l'_i.
\end{equation*}

It means that numbers $l_2, l_3, ..., l_{N}, l'_2, l'_3, ... l'_{N}$ satisfy the conditions of the theorem for $N-1$, which is assumed to be true. Therefore, $l_i = l'_i ~ \forall i > 1$ which is a contradiction. As a result, $l'_1 \neq l_1$; using (A.2) we get $k_1>0$.

Secondly, let's write down condition 2) of the theorem with the equation $l'_i = l_i - k_{i-1} + k_i(l_i+1)$ (which is given by Lemma 1):

\begin{equation*}
    \sum_{i=1}^{N}l_i \geq \sum_{i=1}^N l'_i \implies \sum_{i=1}^N (l'_i-l_i)= \sum_{i=1}^N (-k_{i-1}+k_i(l_i+1)) \leq 0.
\end{equation*}

After expanding it we get the following:

\begin{multline*}
    \sum_{i=1}^N (-k_{i-1}+k_i(l_i+1)) = \sum_{i=1}^N (-k_{i-1}+k_i l_i+ k_i)) = \\
=-k_0 + k_1 l_1 + k_1 - k_1 + k_2 l_2 + k_2 - k_2 + k_3 l_3 + k_3 + ... - k_{N-1} + k_N l_N + k_N = \\
= -k_0 + k_1 l_1 + k_2 l_2 + ... + k_N l_N + k_N \leq 0.
\end{multline*}

Since $k_0=k_N=0$,

\begin{equation*}
    \sum_{i=1}^{N} k_i l_i \leq 0.
\end{equation*}

However, condition 1) states that $l_i>0 ~\forall i$,  Lemma 1 states that  $k_i \geq 0 ~ \forall i > 1$, and we have shown above that $k_1>0$. Therefore $k_i l_i \geq 0 ~ \forall i$ and $k_1 l_1 >0$, which means that $\sum_{i=1}^N k_i l_i > 0$. We get a contradiction which proves the theorem.

\begin{apptheorem}
Let $N$ be some natural number. Let $l_1, l_2, ..., l_N, l'_1, l'_2, ... l'_N$ be natural numbers that satisfy the following conditions: \\
1) $l_i > 0,~ l'_i \geq 0 ~\forall i$;\\
2) $\sum_{i=1}^{N}l_i \geq \sum_{i=1}^N l'_i$;\\
3) $\sum_{i=1}^N l_i Q_i \equiv \sum_{i=1}^N l'_i Q_i \mod{( \sum_{i=1}^{N} l_i Q_i + 1)}$, where $Q_i = \prod_{j=1}^{i-1} (l_j+1)$ (and $Q_1 = 1$). \\
Then $l_i = l'_i ~ \forall i$.
\end{apptheorem}

\emph{Proof.} We will prove this theorem by contradiction. Suppose there are such numbers $l_1, l_2, ..., l_N, l'_1, l'_2, ... l'_N$ that satisfy the conditions of the theorem, but $\exists i : ~ l_i \neq l'_i$. Let's write down condition 3) of the theorem in the following way:

\begin{equation*}
    \sum_{i=1}^N l'_i Q_i = \sum_{i=1}^N l_i Q_i + q (\sum_{i=1}^{N} l_i Q_i + 1),
\end{equation*}

where $q \in \mathbb{Z}$ and $q \geq 0$ since $\sum_{i=1}^N l'_i Q_i > 0$. If $q=0$, than Theorem 1 can be applied and $l_i = l'_i ~ \forall i$, which is a contradiction. From now on we will consider the case $q > 0$. We can rearrange the expression which is multiplied by q:

\begin{multline*}
\sum_{i=1}^{N} l_i Q_i + 1 = 1 + l_1 + l_2(l_1+1) + ... + l_N(l_{N-1} + 1)...(l_1+1) = \\
= (l_1 + 1)(1 + l_2 + l_3(l_2 + 1) + ... + l_N(l_{N-1} + 1)...(l_2+1)) =\\
= (l_1 + 1)(l_2 + 1)( 1 + l_3 + l_4(l_3 + 1) + ... + l_N(l_{N-1} + 1)...(l_3+1)) = \\ = ... = (l_1 + 1)(l_2 + 1)...(l_N+1) = Q_{N+1}.
\end{multline*}

Now condition 3) of the theorem can be rewritten as

\begin{equation*}
    \sum_{i=1}^N l'_i Q_i = \sum_{i=1}^N l_i Q_i + q Q_{N+1}.
\end{equation*}

Let's define $l'_{N+1} \equiv 0$, $l_{N+1} = q > 0$. Then 

\begin{equation}
    \tag{A.4}
    \sum_{i=1}^{N+1} l'_i Q_i = \sum_{i=1}^{N+1} l_i Q_i.
\end{equation}

The numbers $l_1, ..., l_{N+1}; l'_1, ..., l'_{N+1}$ satisfy the conditions of Theorem 1. Condition 1) is satisfied because $l_{N+1} = q > 0$ and $l'_{N+1} \equiv 0$. Since 

\begin{equation*}
\sum_{i=1}^{N+1} l_i > \sum_{i=1}^{N} l_i \geq \sum_{i=1}^{N} l'_i = \sum_{i=1}^{N+1} l'_i,
\end{equation*}

condition 2) of Theorem 1 is also satisfied. Condition 3) is identical to expression (A.4). Therefore, $l_i = l'_i ~\forall i $; but

\begin{equation*}
l_{N+1} = q \neq 0 = l'_{N+1}.
\end{equation*}

This is a contradiction which proves the theorem. 

\end{document}